\documentclass[twocolumn,superscriptaddress,pre]{revtex4}

\usepackage{epsfig}
\usepackage{amssymb,amsfonts,amsmath}
\usepackage[pdftex]{pstricks}

\begin{document}

\title{repgenHMM: a dynamic programming tool to infer the rules of immune receptor generation from sequence data}
\author{Yuval Elhanati}
\affiliation{Laboratoire de physique th\'eorique, UMR8549, CNRS and \'Ecole normale sup\'erieure, 24, rue Lhomond, 75005 Paris, France.}
\author{Quentin Marcou}
\affiliation{Laboratoire de physique th\'eorique, UMR8549, CNRS and \'Ecole normale sup\'erieure, 24, rue Lhomond, 75005 Paris, France.}
\author{Thierry Mora}
\affiliation{Laboratoire de physique statistique, UMR8550, CNRS and \'Ecole normale sup\'erieure, 24, rue Lhomond, 75005 Paris, France.}
\author{Aleksandra M. Walczak}
\affiliation{Laboratoire de physique th\'eorique, UMR8549, CNRS and \'Ecole normale sup\'erieure, 24, rue Lhomond, 75005 Paris, France.}

\date{\today}
\linespread{1}

\begin{abstract}
The diversity of the immune repertoire is initially generated by random rearrangements of the receptor gene during early T and B cell development.
Rearrangement scenarios are composed of random events --~choices of gene templates, base pair deletions and insertions~-- described by probability distributions.  Not all scenarios are equally likely, and the same receptor sequence may be obtained in several different ways.
Quantifying the distribution of these rearrangements is an essential baseline for studying the immune system diversity.
Inferring the properties of the distributions from receptor sequences is a computationally hard problem, requiring enumerating every possible scenario for every sampled receptor sequence.
We present a Hidden Markov model, which accounts for all plausible scenarios that can generate the receptor sequences. We developed and implemented a method based on the Baum-Welch algorithm that can efficiently infer the parameters for the different events of the rearrangement process. We tested our software tool on sequence data for both the alpha and beta chains of the T cell receptor. To test the validity of our algorithm, we also generated synthetic sequences produced by a known model, and confirmed that its parameters could be accurately inferred back from the sequences. The inferred model can be used to generate synthetic sequences, to calculate the probability of generation of any receptor sequence, as well as the theoretical diversity of the repertoire. We estimate this diversity to be $\approx 10^{23}$ for human T cells. The model gives a baseline to investigate the selection and dynamics of immune repertoires.

Source code and sample sequence files are available at \url{https://bitbucket.org/yuvalel/repgenhmm/downloads}.
\end{abstract}

\maketitle

\section{Introduction}

The ability of the adaptive immune system to identify a wide range of threats rests upon the diversity of its lymphocyte receptors, which together make up the immune repertoire. Each such receptor can bind specifically to antigenic molecules, and initiates an immune response against the threat.

T cell receptors (TCR) are composed of two protein chains, called alpha and beta. B cell receptors (BCR) share a very similar structure, with a light chain and heavy chain playing the same role. Each chain is produced according to the same process of V(D)J rearrangement. In each new cell and for each of the two chains, two germline segments for alpha chains (V$\alpha$ and J$\alpha$ genes), or three segments for beta chains (V$\beta$, D$\beta$ and J$\beta$ genes), are assembled together to form the recombined gene coding for the chain. In addition, at the junctions where the segments are joined, the ends of the segments are trimmed, and random nucleotides are inserted (see Fig.~\ref{diagram}a for a diagram describing the alpha chain rearrangement process). 
This process creates a large initial diversity of possible receptors, which are later selected according to their recognition functionality.
An important property of this process is that it is redundant, as many different V(D)J rearrangements may lead to the exact same sequence. It is thus impossible to unambigously reconstruct the scenario from the sequence alone, a problem that is aggravated by sequencing errors.

The rearrangement process is random, as is each of the elements composing it --~choice of germline segments, number of deleted nucleotides, number and identity of insertions.

Since the rearrangement process is the basis of repertoire diversity, it is  important to study its distribution quantitatively. With recent advances in high throughput sequencing, there is a growing body of data on repertoires for both T and B cells, in a variety of situations. Using large sequence datasets of rearranged, non-productive genes, the probability distribution of rearrangement events in human TCR beta chain and BCR heavy chains could be inferred using statistical methods, gaining important insights into the random processes underlying repertoire diversity \citep{Murugan2012,Elhanati2015}. However, these studies are based on a brute force approach, which enumerates every possible rearrangement scenario for each observed sequence. This is a very computationally costly procedure, which is unrealistic for very large datasets. 

In this report we present a dynamic programming approach to learn the distribution of rearrangement scenarios from large numbers of non-productive sequences in an efficient way. This approach is based on a Hidden Markov Models (HMM) formulation of the problem, and learns its parameters using a modified Baum-Welch (BW) algorithm to avoid the full enumerations of all scenarios. 
Many studies have described algorithms designed to process large numbers of rearranged TCR or BCR genes and extract the template V(D)J genes of the rearrangement \citep{Paciello2015,Bonissone2015,Russ2015,Nazarov2015,Brochet2008,Ye2013,Souto-Carneiro2004,Ohm-Laursen2006,Gadala-Maria2015,Frost2015,gaeta2007,munshaw2010,Wang2008,Volpe2006a,Ralph2015}. These tools process each sequence separately to obtain the best (but often incorrect) alignment to a V(D)J combination, sometimes using dynamic programming or HMM \citep{gaeta2007,munshaw2010,Volpe2006a,Ralph2015}, and assume an implicit, {\em ad hoc} prior on rearrangements. By contrast, our algorithm explores all plausible alignments for each sequence from data to learn accurately the distribution of rearrangement events.

Once the model of rearrangement has been learned by our procedure, the entire distribution of possible sequences and their probabilities is accessible. Our algorithm can calculate the probability of any sampled sequence, even if it is not part of the data used to learn the model, and it can generate arbitrary numbers of synthetic sequences with the exact same statistics as the data. It can also calculate the entropy of the rearrangement process --~a classical measure of sequence diversity. This enables us to further our understanding of the generation process, quantify the baseline state of the immune system and evaluate subsequent processes such as somatic selection. Finally, our work produces insights not just on the data sequences, but on the underlying biological processes.

\section{Methods}
\subsection{Model}
The algorithm assumes a general form for the probability distribution of possible rearrangements, and then finds the parameters of that distribution that best fit the sequence data \citep{Murugan2012,Elhanati2015}. For simplicity we first describe the model for the alpha chain of TCRs, which also applies for the light chain of BCRs. The case of the beta and heavy chains will be described later.

The model specifies probability distributions for each of the rearrangement events: V and J gene choices $P(V,J)$, number of deletions conditioned on the gene being deleted $P(\text{del}V|V)$ and $P(\text{del}J|J)$, and insertion length and nucleotide identity $P(\rm{ins})$. Together these distributions form the parameter set $\theta=\{P(V,J),P(\text{del}V|V),\ldots\}$.
The probability of a given rearrangement scenario $r=(V,J,{\rm del}V,{\rm del}J,{\rm ins})$ is then given by:
\begin{equation}
P_\text{rearr}(r|\theta) = P(V,J) P(\text{del}V|V) P(\text{del}J|J) P({\rm ins}).
\label{model_eq}
\end{equation}

The specific form of the model assumes some dependencies between the events. In the above formula, for instance, the probabilities of each V and J choice can be dependent, but the insertion is independent from both, while the deletion probabilities are dependent only on the identity of the gene being deleted. 
The model choice is done based on biological knowledge, and has been validated by previous work in the case of the beta chain \citep{Murugan2012}.

To avoid confounding factors related to thymic or peripheral selection, the inference of the parameters is performed on non-productive genes. During the maturation of cells, some rearrangement events produce non-productive genes that are either out of frame, having the wrong combination of insertions and deletions, or contain a stop codon.

When this happens, the other chromosome in the same cell may undergo a second successful rearrangement event, ensuring the survival of the cell.
Yet the non-productive rearrangements remain and are part of the sequence dataset. Since these sequences have no effect on the cell in which they reside, they have not been selected.  Studying their statistics is thus equivalent to studying the generation process itself, with no selection.

While the model specifies the distribution over rearrangement scenarios, the data consists of recombined sequences, denoted by ${\bf s}$, which can be the result of different scenarios $r$. 
The recombination events are hidden variables, and the likelihood of a sequence is the sum of the probabilities of all scenarios leading to it, $P({\bf s})=\sum_{r\to {\bf s}}P_{\rm rearr}(r)$.
The likelihood of the sequence dataset cannot easily be maximised with respect to the model parameters.
To overcome this problem, the Expectation-Maximisation (EM) algorithm can be used to maximise the likelihood in an iterative manner.

In each iteration, new model parameters $\theta'$ are chosen to increase the likelihood until the maximum is obtained. In the expectation step, the log-likelihood of hidden rearrangements is averaged over their posterior distribution under the current model $\theta$, to form
$Q(\theta'|\theta)=\sum_{a=1}^n \sum_{r} P(r|{\bf s}^{(a)},\theta) \log P_{\rm arran}(r|\theta')$,
where the sum on $a$ runs over the $n$ sequences $({\bf s}^{(1)},{\bf s}^{(2)},\ldots,{\bf s}^{(n)})$ in the dataset. The maximisation step consists of maximising $Q(\theta'|\theta)$ over $\theta'$ to obtain the new parameter set. Because of the simple factorised form of Eq.~\ref{model_eq}, this second step is equivalent to calculating the frequency of each rearrangement event under the posterior $P(r|{\bf s}^{(a)},\theta)=P(r,{\bf s}^{(a)}|\theta)/P({\bf s}^{(a)}|\theta)$. For example, $P(V,J)$ is updated according to:
\begin{equation}
P'(V,J)=\frac{1}{n}\sum_{a=1}^n \sum_{r:V,J} P(r|s^{(a)},\theta),
\end{equation}
where the sum on $r:V,J$ runs over scenarios with gene choices $V$ and $J$. Similar update rules are used for the other model parameters $P({\rm del}V|V)$, $P({\rm del}J|J)$, and $P({\rm ins})$. The sums over possible scenarios, for each data sequence $s^{(a)}$, are computationally heavy. To make them easier, we now introduce an equivalent HMM formulation of the model.

\subsection{HMM formulation}

The almost linear structure of rearrangements allows for their description as a Markov chain.
Hidden Markov models lend themselves to the much more efficient forward-backward algorithm for marginal estimations, and in combination with Expectation-Maximisation, the Baum-Welch (BW) algorithm, for parameter inference. In general however, the V and J gene choices may be correlated in their joint distribution $P(V,J)$, breaking the Markovian nature of the rearrangement statistics. To preserve the Markovian structure, we built a separate HMM for each choice of the pair of germline genes $(V,J)$, and use the  forward-backward algorithm to calculate the marginals of the other rearrangement events conditioned on that choice. These conditional marginals will then be combined for all $(V,J)$ to perform the maximisation step of EM.

\begin{figure}

\centering

\includegraphics[width=\linewidth]{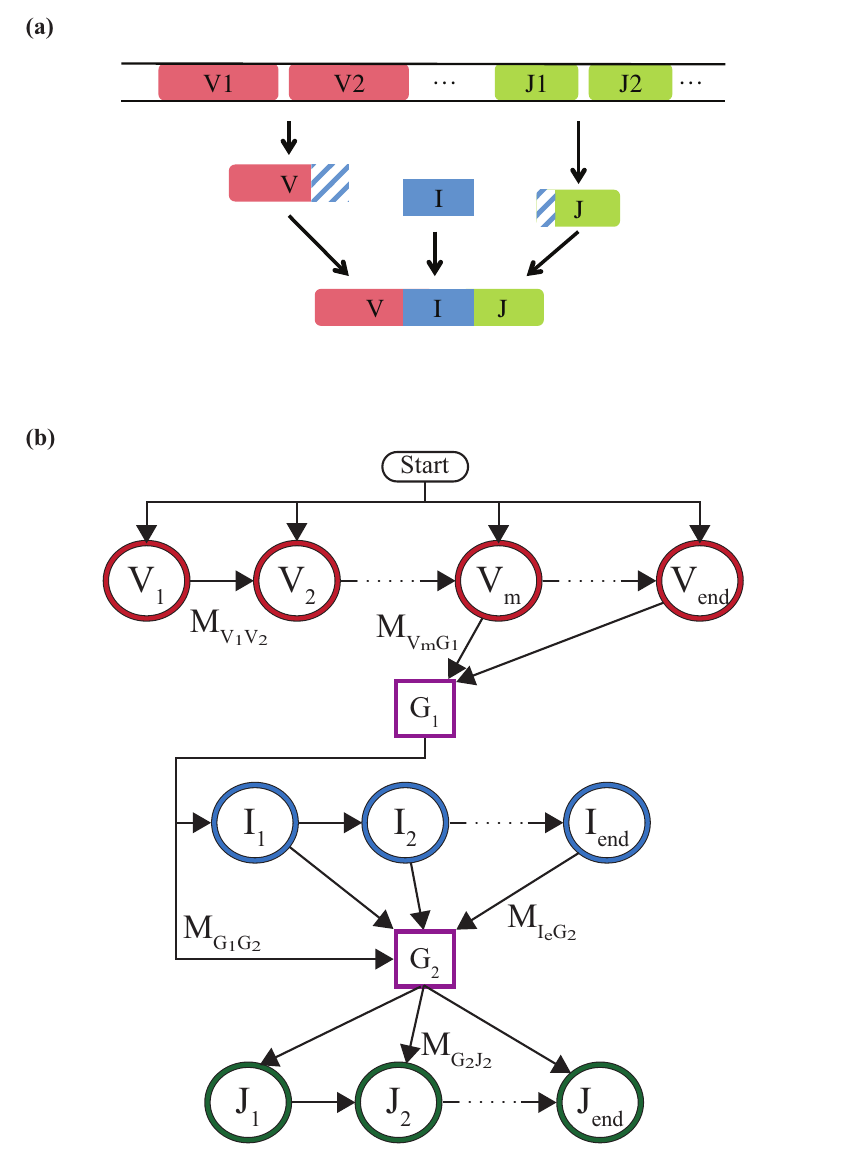}
\caption{(a) Schematic description of the rearrangement process for the light and alpha chains. Random V and J genes are chosen from the genome. A random number of nucleotides are trimmed from their facing ends. These ends are then joined with an insertion segment of variable length and composition. (b) Markov model for this rearrangement process, when the V and J gene choices are known. By progressing one path following the arrows, the model produces a rearranged receptor gene. Each state denoted by a circle emits a nucleotide. V and J states each emit one nucleotide from the chosen template, up to an error rate. Emissions from the I states are drawn from an specified distribution. The states represented by squares are nonemitting ghost states. The arrows represent the allowed transitions, some of them are marked on the diagram with $M_{SS'}$. The probabilities of the transitions and emissions are the parameters of the HMM, as described in the main text.}
\label{diagram}
\end{figure}

For a chosen $(V,J)$ pair, a Hidden Markov Model (HMM) is constructed to yield the recombined sequences in accordance with Eq. \ref{model_eq}. Fig.~\ref{diagram}b shows a diagram of the model. The model proceeds through a sequence of hidden states $S$, which emit nucleotides $s_i$, for $i=1,\ldots,L$ where $L$ is the sequence length, thus producing the entire sequence ${\bf s}=(s_1,\ldots,s_L)$.

We distinguish between two types of emitting states, represented by circles in Fig.~\ref{diagram}b. First, `genomic states,' or V and J states, are defined for each position on the genomic templates $V$ and $J$, and are denoted by $V_1,V_2,...,V_\text{end}$ for V states, and likewise for J states.
These states emit the nucleotide encoded in the genomic template at the corresponding position, or, with a small error probability $p_\text{err}$, a different nucleotide. These non-templated emissions can be caused by sequencing errors, B cell hypermutations or uncharted alleles.
Second, `insertions states' emit random nontemplated nucleotides according to a distribution. $I_1$ corresponds to the first inserted nucleotide, $I_2$ to the second one, and so on.
In addition, we introduce two `ghost states' $G_1$ and $G_2$, represented by rectangles in Fig.~\ref{diagram}b, between the V and I states, and between the I and J states. These states do not emit nucleotides and their sole function is to reduce the number of possible transitions between states by isolating the state types, thus easing computations.

Each sequence is the result of a stochastic path through a series of states, defined in Fig.~\ref{diagram}b,  and their random emissions.
To illustrate how the HMM operates, we follow a possible path leading to the production of a light chain for a given choice of V and J. The chain starts from the $V_1$ state, going along the V gene and emitting nucleotides from the gene. Most of these nucleotides are those of the genomic template, up to the error rate. At some point (possibly before all V states are exhausted to account for potential V deletions) the process transitions to the first ghost state $G_1$. From $G_1$ the process goes to the first insertion state $I_1$, or directly to $G_2$ if there are no insertions. Each
insertion state emits a nucleotide. After a certain number of insertions, the process moves to the second ghost state, $G_2$, and then on to a J state (but not necessarily $J_1$ to account for J deletions). Finally the process will continue along the J states until $J_{\rm end}$, completing the sequence.

The HMM is defined by two sets of parameters which map directly onto $\theta\backslash P(V,J)$. The first set of parameters is the transition probabilities $M_{SS'}$ between any two states $S$ and $S'$ connected by an arrow in Fig.~\ref{diagram}. The transition rates between V states and $G_1$, and between $G_2$ and J states, can be mapped onto the deletion profiles of the V and J genes respectively, and the transition rates between the I states and $G_2$ can be mapped onto the distribution of the number of insertions. 
The transition matrix $M_{SS'}$ is very sparse, thanks in part to the ghost states,

allowing for quick computations as we will see below.
The second set of parameters are the emission probabilities $E_S(s)$ that nucleotide $s$ is emitted by state $S$. If $S$ is a genomic (V or J) state, then $E_S(s)=1-p_{\rm err}$ if $s$ is the template nucleotide, which we denote by $\sigma_S$, and $p_{\rm err}/3$ otherwise. If $S$ is an insertion state, it is given by a distribution $E_{I}(s)$, which we assume to be common to all insertion states, {\em i.e.} independent of the order of insertions.

\subsection{A modified Baum-Welch algorithm}

The Baum-Welch (BW) algorithm finds the parameters for a given HMM which maximise the likelihood of producing the observed sequences \citep{durbin1998,bishop2006}. It is an instance of the EM algorithm, where the maximisation step is performed using the forward-backward algorithm.
Since our HMM is conditioned on the knowledge of the $(V,J)$ pair, which is itself a hidden variable, BW cannot be applied without modification.
However, we can still use the forward-backward algorithm to calculate the posterior probabilities of rearrangement events for a given sequence ${\bf s}=(s_1,\ldots,s_L)$ and a given putative $(V,J)$ choice, and combine these probabilities at the end.

The forward pass of the forward-backward algorithm calculates $\alpha_i(S)$, the joint probability of the model being in a specific state $S$ and emitting the sequence up to the $i^{\rm th}$ nucleotide, ($s_1,...,s_i$). The backwards pass does the same for $\beta_i(S)$, the conditional probability of emitting the sequence upstream from position $i$, given that the state in this position is $S$:
\begin{eqnarray}
\alpha_i(S)& :=& P(s_1, ..., s_i, S|V,J), \\
\beta_i(S) &:= &P(s_{i+1}, ..., s_L | S, V,J).
\end{eqnarray}
These probabilities are calculated using the following recursion relations:
\begin{eqnarray}
\alpha_i(S)& =& E_S(s_i) \sum_{S'} {M_{SS'} \alpha_{i-1}(S') },\\
\beta_i(S) &= & \sum_{S'} {E_{S'}(s_{i+1}) M_{S'S} \beta_{i+1}(S')}.
\end{eqnarray}
Since our transition matrix $M_{SS'}$ is very sparse, the sum over $S'$ has few terms and can be calculated efficiently. Having obtained these forward and backward probabilities for a sequence given a choice of V and J genes, the posterior marginal probabilities for each transition ($S\to S'$), as well as the posterior emission probabilities are calculated as
\begin{eqnarray}
P(S\to S'|V,J,{\bf s}) & \propto &\sum_i {\alpha_i(S)M_{S'S}E_{S'}(s_{i+1})\beta_{i+1}(S')},  \nonumber \\
P({\text{err}}|V,J,{\bf s})& \propto &\sum_i {\sum_{S \in V,J} {\alpha_i(S) \beta_n(S)}}(1-\delta_{\sigma_S,s_i}),\nonumber  \\
P_{\text{ins}}(s|V,J,{\bf s})& \propto & \sum_i {\sum_{S \in I} {\alpha_n(S) \beta_n(S)}}\delta_{s,s_i},
\label{VJ_spec_marg}
\end{eqnarray}
up to a normalisation constant, where $\delta$ denotes Kroeneker's delta.

The existence of ghost states requires making a small adjustment to this scheme. 
Each ghost state introduces an offset between the state index and the corresponding position on the sequence. Thus, V states in the $n$ position correspond to position $n$ in the sequence, I states to position $n-1$ in the sequence, and J states to position $n-2$.

Once the posterior marginals have been evaluated for each data sequence, and for each putative choice of $V$ and $J$, we can combine them to obtain
the update equations of our modified Baum-Welch algorithm:
\begin{eqnarray}
M_{S'S}&\leftarrow& \frac{1}{n} \sum_{a=1}^n \sum_{V,J} P(V,J|{\bf s}^{(a)}) P(S\to S'|V,J,{\bf s}^{(a)}),\nonumber\\
p_{\rm err}&\leftarrow& \frac{1}{n} \sum_{a=1}^n \sum_{V,J} P(V,J|{\bf s}^{(a)}) P({\text{err}}|V,J,{\bf s}^{(a)}),\\
E_{S}(s)&\leftarrow& \frac{1}{n} \sum_{a=1}^n \sum_{V,J} P(V,J|{\bf s}^{(a)})P_{\text{ins}}(s|V,J,{\bf s}^{(a)}),\nonumber
\end{eqnarray}
where the posterior probability $P(V,J|{\bf s}^{(a)})$ is calculated using Bayes' rule, $\propto P({\bf s}^{(a)}|V,J) P(V,J)$, with $P({\bf s}^{(a)}|V,J)=\sum_S \alpha_L(S)$.

Finally, the algorithm outputs the probability that any sequence ${\bf s}$ was generated as $P({\bf s})=P({\bf s}|V,J)P(V,J)$. This probability can be used to calculate the log-likelihood of the model, $\sum_{a=1}^n \log P({\bf s}^{(a)})$, and track the progress of the BW algorithm at each iteration.

\subsection{Pre-Alignments}
For a given sequence, there may be many potential candidates for the segments $(V,J)$, but not all are equally plausible, especially when sequence reads are long, and not all should be considered.
Before starting the inference procedure, the sequences are locally aligned against all genomic templates using the Smith-Waterman algorithm. By creating a shortlist of genomic genes that had an alignment score larger than a tunable threshold, the inference procedure can exclude certain gene choices \textit{a priori}. This saves considerable computation time by omitting rearrangement scenarios that would have a negligible effect 
due to high numbers of errors.

In addition, this pre-alignment procedure provides us with a mapping between the positions along the sequence read and each genomic gene. Thus, each of the V and J states of the HMM may only be present at a single position along the sequence, drastically limiting the number of states that we need to consider at each position and improving the speed of computations.

\subsection{Beta and heavy chains}

\begin{figure}

\centering

\includegraphics[width=\linewidth]{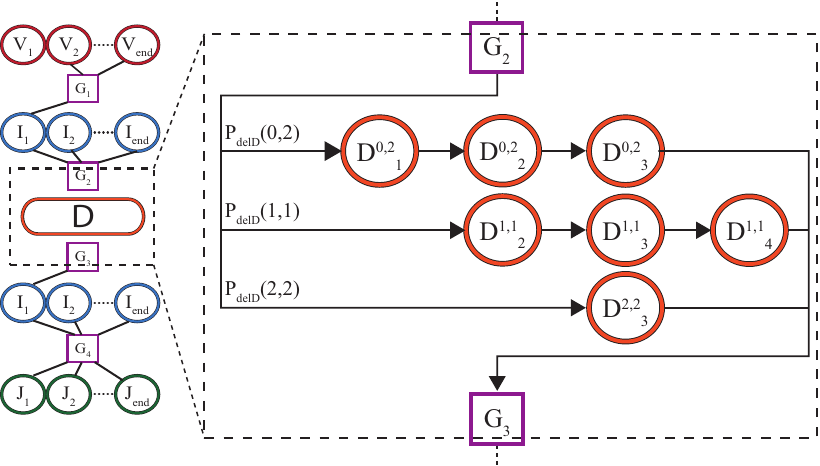}
\caption{Subdiagram of Markov model for beta and heavy chain, focusing on the D gene. Each row corresponds to a different pattern of deletions $({\rm del}Dl,{\rm del}Dr)$ for the left and right ends of the D segments.
State $D^{({\rm del}Dl,{\rm del}Dr)}_d$ corresponds to the $d$th base in the D gene, when $l$ bases are deleted from the left and $r$ from the right. Each row is entered from the ghost state $G_2$ with probability $P_{{\rm del}D}(l,r)=P({\rm del}Dl,{\rm del}Dr)$, and then proceeds deterministically until $G_3$.}
\label{Ddiagram}
\end{figure}

For the beta chain of the TCR (or the heavy chain of the BCR), the model is similar to the one in Eq.~\ref{model_eq}, with the addition of a D gene choice, its deletions from both the left and right sides ($\text{del}Dl,\text{del}Dr$) and two independent insertion events at the VD and DJ junctions ($\text{ins}VD,\text{ins}DJ$):
\begin{equation}
\begin{split}
& P_\text{rearr}(r|\theta) = P(V,D,J) P(\text{del}V|V) P(\text{ins}VD)  \\
&\quad \times P(\text{del}Dl,\text{del}Dr|D) P(\text{ins}DJ) P(\text{del}J|J).
\end{split}
\end{equation}
A similar HMM as for the alpha chain can be built by adding genomic D states and having two types of insertion states, for the VD and DJ junctions, and extra ghost states to separate them. Then, the procedure described above can be applied {\em mutatis mutandis}.

In addition to the V and J gene, the D gene has to be chosen. An HMM is built for each triplet of genomic segments $(V,D,J)$.
D genes are short and deleted on both sides. For this reason, they are much harder to align. Even the position of a candidate genomic D segment along the sequence is no longer unambiguous as is the case for V and J segments.
For this reason, we do not pre-align the D genes to the sequence. Instead, for each sequence all D genes with all possible locations are considered. Technically this means that the sequence positions at which D states may occur are not pre-specified.

During the rearrangement process D genes are deleted from both sides. The number of bases truncated from the left and right ends of the gene are correlated --~in the extreme case, the sum of both deletions cannot exceed the length of the gene. Since the left and right D deletions correspond to transitions from non-adjacent states, this correlation cannot be described using a Markov model. We have addressed this issue by enumerating all the possible deletions from the left and the right as separate states. In practice, we define different types of D states for each choice of deletions, as depicted in Fig.~\ref{Ddiagram}. Each D deletion profile $(\text{del}Dl,\text{del}Dr)$ defines a separate subchain of the Markov chain going along the D gene, which is entered from the previous ghost state with probability $P(\text{del}Dl,\text{del}Dr|D)$.

\subsection{Entropy estimates}
\label{sec_ent}
The inferred model can be used to characterise the diversity of the distribution of all possible receptors, and not just of the sampled sequences used when inferring that distribution. We quantify the diversity of a stochastic quantity $X$ using the Shannon Entropy: $H(X)=-\sum_{X}{P(X)\log_2 P(X)}$ (measured in bits). For instance, $H({\bf s})$ gives a measure of sequence diversity. 
Since a uniform distribution with $2^H$ outcomes has entropy $H$, the number $2^H$ can, even for non-uniform distributions, be interpreted as an effective diversity number, sometimes called true diversity.
The entropy of rearrangements can be calculated explicitly thanks to the factorized form of the distribution. For example, for alpha chains:
\begin{equation}
\begin{split}
H(r) & = H(V,J) +  \sum_V P(V) {H({\text{del}V|V})}  \\
& +\sum_J P(J){H({\text{del}J|J})} + H(\text{ins}).
\end{split}
\label{eq_ent}
\end{equation}
This expression clearly separates the contributions from $(V,J)$ segment choice, deletions and insertions.
The entropy of sequences $H({\bf s})$ cannot be calculated in this way since receptor sequences can be produced by multiple rearrangements, but can easily be estimated by averaging $\log P({\bf s})$, where $P({\bf s})$ is calculated by the forward-backward algorithm as explained above.

\section{Results}

\subsection{Implementation}
The method was implemented in C++ (std11) as a command line tool. OpenMP was used for parallelization. 

The main pipeline has two parts, the alignment module and the inference module. The input for the entire pipeline is a FASTA or plain text file with unique recombined and non-productive nucleotide sequences, and FASTA files with the genomic templates for the different V, J germline segments (as well as D in the case of heavy or beta chains). 
Genomic files can be obtained from genomic databases such as IMGT \citep{Giudicelli2005}. The alignment code was written to read IMGT FASTA files and extract gene names.

Between the alignment and inference procedures, alignments below a certain threshold are discarded to improve performance. The threshold can be tuned for different data sets. Sequences that do not align well to at least one known gene of each type are completely eliminated as a quality control. Setting the thresholds should be done carefully, keeping a large majority of the sequences with at least one good alignment, but excluding ones which had only low score alignments. To this end, an auxiliary module is included that outputs statistics on the best alignment score for each sequence. Curated alignment files are saved at the end of the alignments stage, and used as input for the inference. Apart from the alignments, the only parameters needed for inference are the maximum numbers of insertions and deletions.

The output of the main pipeline is the value of the model parameters: the joint distribution of segment usage; the distributions of deletions at each of the deletion sites conditioned on the gene; the distribution of the number and composition of inserted nucleotides at the junctions; and a global error rate accounting for sequencing errors, hypermutations or genomic variants.

Two more modules are included. First, given a list of sequences and an inferred model, the software can compute generation probabilities for all sequences. Second, a synthetic sequence generation module that can produce sequences from a given model. This module can be used to study the properties of the distribution or to verify the inference algorithm using a known model, as we will see below.

\subsection{Application to alpha chain data}
\begin{figure}
\begin{center}
\noindent
\includegraphics[width=\linewidth]{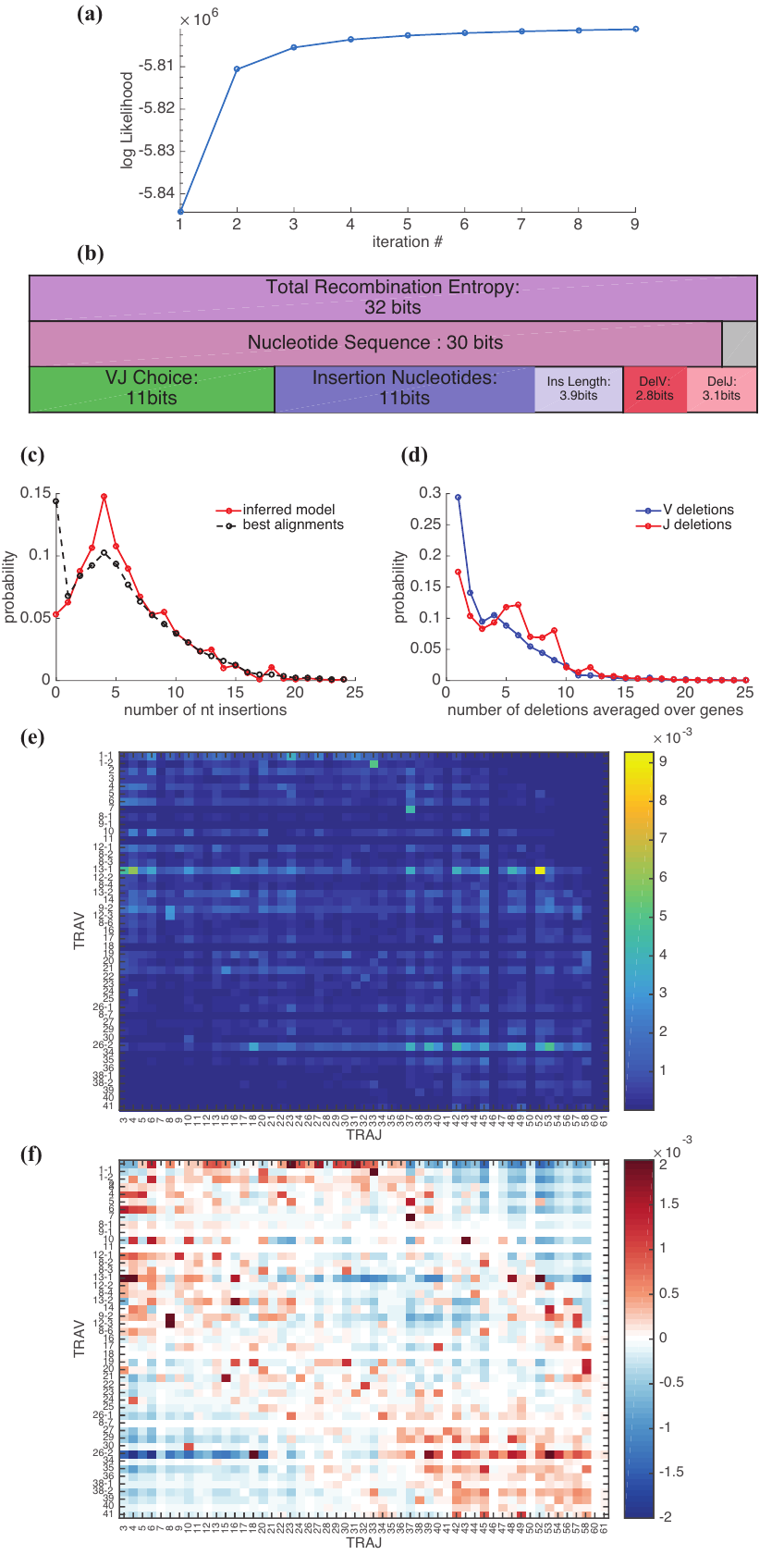}

\caption{TCR alpha chain rearrangement distribution inferred from sequence data taken from \citep{Zvyagin2014}. (a) The log-likelihood of the data given the model saturates as a function of the number of iterations of the Expectation-Maximisation algorithm. (b) Shannon entropy of rearrangements (top row) and sequences (middle row). The sequence entropy is lower than the total recombination entropy because of convergent rearrangements.
The rearrangement entropy is the sum of entropies of its elementary events (bottom row). (c) Distribution of the number of inserted nucleotides (red curve). For comparison, the same distribution obtained by taking the best alignment of each sequence to the genomic templates is represented by the black dashed line.
(d) Distributions of the number of deletions for both V and J genes. These distributions, which are gene dependent, are here averaged over genes.
(e) Joint distribution for V and J usage, $P(V,J)$. Genes are ordered by position along the genome. (f) The covariance $P(V,J)-P(V)P(J)$ clearly shows strong correlations for genes that are either close to the separation between the V and J segments, or far from it.}
\label{alpha_results}
\end{center}
\end{figure}

The algorithm was applied to human TCR alpha chain sequences from \cite{Zvyagin2014}. The data consist of around 80,000 non-productive sequences, each 151bp long, covering both the V and J segments. Sequences were aligned to lists of genomic sequences from the IMGT online database \citep{Giudicelli2005}, and then given as input to the inference algorithm. The model converged rapidly as can be seen by the quick saturation of the likelihood (Fig. \ref{alpha_results}a).

The entropy of the rearrangement process quantifies the diversity of possible scenarios. It was calculated using Eq. \ref{eq_ent} and found to be 32 bits (Fig.~\ref{alpha_results}b, top line). This entropy can be partitioned into contributions from each of the rearrangement events --~segment choice, insertions and deletions (bottom line). The largest contribution to the entropy comes from the insertions, followed closely by gene choice. We also estimated the entropy of the sequence distribution (middle line), which is smaller than the rearrangement entropy because of convergent rearrangements --~multiple rearrangements leading to the same sequence (grey box). This estimate was based on samples of simulated sequences. Undersampling bias and error were corrected for by using samples of different sizes and validating the convergence of the entropy. The entropy of the alpha chain sequences is 30bits, which corresponds to a diversity number of about $10^9$. 

Inferred insertion and deletion profiles for the alpha rearrangements are shown in Figures~\ref{alpha_results}c and d, with the deletion profile averaged over genes. The peak of the insertion distribution is at 5bp, similar to previous results for the beta chain.
The joint distribution of gene usage for V and J, represented in Fig.~\ref{alpha_results}e, shows a wide variety of gene usage probabilities, with clear dependencies on the ordering of genes according to their location on the chromosome \citep{lefranc2001t}.
To better see these dependencies, in Fig.~\ref{alpha_results}f we plotted $P(V,J)-P(V)P(J)$ as a measure of the correlation between V and J choices. 

In \cite{Warmflash2006}, it was proposed that rearrangements can occur in several steps. When a V and a J segment are joined, the genomic segments that were between them are excised, but the segments located on the outer flanks remain.
Then, successive rearrangements joining these outer segments might occur. It was hypothesised that early joining events involve V and J genes that are close to each other, hence proximal to the boundary between the V and J cassettes. Later joining events, on the other hand, should involve more distal genes as the proximal genes are likely to have been excised.
This phenomenon is expected to lead to correlations between pairs of genes which are either both distal or both proximal, which is consistent with the results of Fig.~\ref{alpha_results}f. Notice also that in the intermediate range our model predicts low correlation within a certain window of distances between the V and J genes.

\begin{figure}
\begin{center}
\noindent
\includegraphics[width=\linewidth]{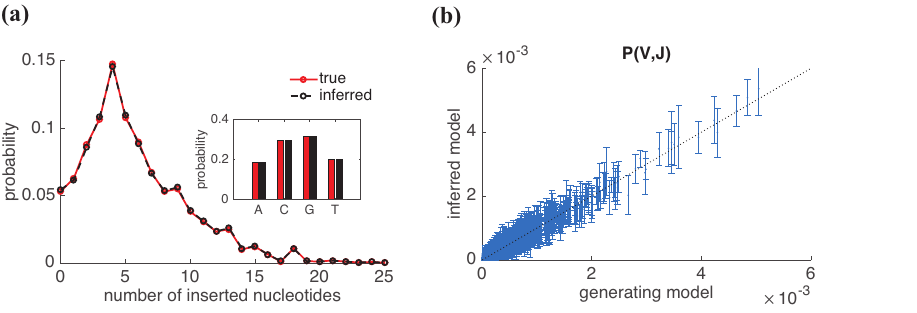}

\caption{Performance of the algorithm on synthetic data. Sequences generated using a known model were given as an input to the inference algorithm. The results of the inference are compared to the true model used for generation, for (a) the distribution of the number of insertions (inset: usage of inserted nucleotides), and (b) V, J gene usage. The error bars, which correspond to sample noise, are smaller than symbol size for (a).}
\label{alpha_simulated}
\end{center}
\end{figure}

\subsection{Test on synthetic data}
In order to check the validity of the algorithm, we ran it on sequences that were produced according to a known model. We generated 100,000 synthetic sequences according to the model learned in the previous section, and relearned a model from these sequences using our algorithm. In Fig.~\ref{alpha_simulated} we compare the parameters of the model used for generation to those of the inferred model. Sampling was repeated 5 times to estimate sample noise, which was found to be very small for all paramaters, except for gene usage (error bars in Fig.~\ref{alpha_simulated}b).

The insertion bias, {\em i.e.} the usage of the different nucleotides in inserted segments, is inferred almost perfectly, as seen in Fig.~\ref{alpha_simulated}a. The probabilities for each V,J choice also show excellent agreement, within sampling errors (Fig.~\ref{alpha_simulated}b).
The distribution of the number of insertions also agrees very well (Fig.~\ref{alpha_simulated}c). However, when inferring the same model, but with a higher error rate ($1\%$), we found that the inferred insertion distribution noticeably overestimated the probability of zero insertions.

\begin{figure}
\begin{center}
\noindent 
\includegraphics[width=\linewidth]{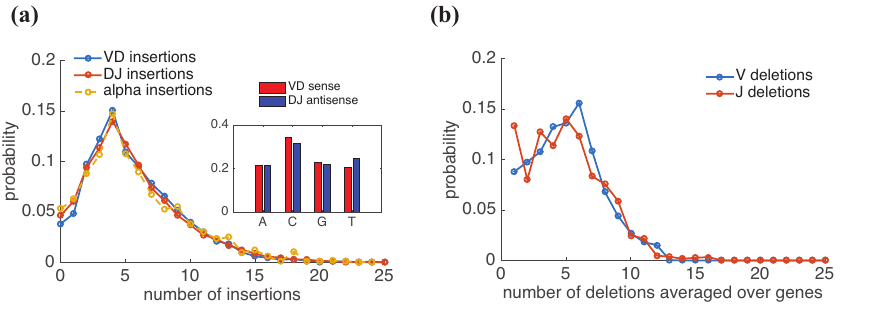}

\caption{TCR beta chain rearrangement distribution inferred from sequence data previously analysed in \citep{Murugan2012}.
 (a) Distribution of the number of insertions at both VD and DJ junctions, and comparison with the distribution of insertion in the alpha chain from Fig.~\ref{alpha_results}c. Inset: The nucleotide usage is identical for VD and DJ insertions when considered on opposite strands. (c) Distribution of the number of deletions on both the V and J genes, averaged over different genes.}
\label{beta_results}
\end{center}
\end{figure}

\subsection{Application to beta chain data}

The beta and heavy chain algorithm was applied to the human TCR beta data that was already analysed in \citep{Murugan2012} using brute-force methods. The inferred model parameters were all very similar to those reported in \cite{Murugan2012},  confirming the validity of our algorithm. The distribution of the number of insertions and deletions are displayed in Fig.~ \ref{beta_results}. Remarkably, insertion profiles at the two insertion sites are very close to each other, as previously reported, but they also closely match the insertion profile of the alpha chain (Fig.~\ref{beta_results}a). Nucleotide usage in each of the two  inserted regions (between V and D, and between D and J) is shown in the inset. The VD insertion base usage is similar to the usage of the complementary bases (antisense) in the DJ region, suggesting that the biological mechanism is operating on the opposite strands for both insertions types, as previously noted \citep{Murugan2012}.
From the computational point of view, because the algorithm enumerates both the D gene choice and its deletions, its benefit is smaller for beta chains than for alpha chains.

\section{Discussion}
The strength of the adaptive immune system lies in its inherent diversity. This dynamic diversity is the result of several stochastic biological mechanisms, including complex enzymatic reactions that alter the DNA structure and composition. The action of some of the enzymes such as RAG and TdT have been studied extensively (see \cite{Schatz2011a}, \cite{Jung2004} and \cite{Schatz2011} for reviews).
In our work we have studied the way in which the diversity is generated in a top down approach, focusing on statistical properties as inferred from sequenced receptor genes. In principle, this is a computationally difficult problem, due to the very large number of possible rearrangement scenarios. 

The algorithm described here can be used to study the properties of generation of receptor chains of B and T cells in any species, from large sequence data sets. Our dynamic programming approach, which is a variant of the Baum-Welch algorithm, takes advantage of the linear structure of rearrangements to avoid a full enumeration of scenarios. 
In a brute-force approach such as the one presented in \cite{Murugan2012} and \cite{Elhanati2015}, the algorithmic complexity scales as the product of the numbers of each independent rearrangement event.
By contrast, the complexity of the method we presented here is additive with the number of insertions and deletions.

Despite technological advances, sequencing techniques still introduce errors. In addition, allelic variants and hypermutations in B-cells introduce additional discrepancies between known template genes and sequence reads. Our method can be used with models that account for such events. In the presented version of the algorithm, substitution errors are already fully accounted for.
Insertion and deletion errors could likewise be handled by adding transitions that skip or repeat bases.

We find many common features of generation for the alpha and beta chains of the TCR. There is a difference of diversity, due to the greater length and complexity of the beta compared to the alpha chain. The diversity number of the beta chain repertoire, estimated to be approximately $10^{14}$ in \cite{Murugan2012}, is therefore much larger than that of the alpha chain reported here, $10^9$. Assuming that the rearrangements of the alpha and beta chains are independent, the total TCR diversity is about $10^{23}$. 

Inferring statistical properties of the underlying biological processes can be thought of as a diagnostic tool for the properties of the immune system, and could be used in a variety of clinical settings. The generation process and subsequent selection shape the initial diversity of the immune system, and we have found this process to be remarkably universal across normal, healthy humans, expect for slight variations in gene usage (\cite{Murugan2012,Elhanati2015}). However, infections or irregularities in the immune system can be seen as perturbations that will change these distributions. By comparing the normal distributions to different pathological situations, information on the reaction of the immune system can be extracted. This information could in turn be used for diagnosis, using fast computational tools such as the one presented here.

With more and more immune repertoire sequence data being collected, efficient algorithms are needed. The ability to quickly infer and analyse large data sets is essential both for our basic understanding of the adaptive immune system and also for specific clinical applications. 

\bigskip

\section*{Acknowledgements}
We thank members of the Chudakov and Lebedev groups at the Institute of
Bioorganic Chemistry of the Russian Academy of Sciences, and in particular M. Pogorelyy, for sharing their TCR alpha data with us. This work was supported by European Research Council Starting Grant n. 306312.

\bibliographystyle{pnas}

\bibliography{refs}
\end{document}